\documentclass[11pt]{article}

\usepackage[utf8]{inputenc}

\usepackage{cite}

\usepackage{nameref,hyperref}
\usepackage{eucal}
\usepackage{amsmath,amssymb}

\usepackage[right]{lineno}

\usepackage{microtype}

\usepackage{changepage}

\usepackage[aboveskip=1pt,labelfont=bf,labelsep=period,singlelinecheck=off]{caption}

\usepackage{lastpage,fancyhdr,graphicx}
\usepackage{epstopdf}

\usepackage{color}

\definecolor{Gray}{gray}{.25}

\usepackage{graphicx}

\usepackage{sidecap}

\usepackage{wrapfig}
\usepackage[pscoord]{eso-pic}
\usepackage[fulladjust]{marginnote}

\def\w{\omega}

\def\Re{\mathcal{R}\mathrm{e}}
\def\R{\mathbb{R}}

\begin{document}


\author{
		\textbf{Sami Karkar $^1$, Emanuele De Bono $^1$ and Manuel Collet $^1$} \\
		$^1$ LTDS, \'Ecole Centrale de Lyon, CNRS UMR5513\\
		36 avenue Guy de Collongue, 69134 Ecully Cedex (France) \\
		\medskip
		e-mail: \textbf{sami.karkar@ec-lyon.fr}, \textbf{emanuele.de-bono@ec-lyon.fr}\\	
		\textbf{Gaël Matten $^2$, Morvan Ouisse $^2$} \\
		$^2$ FEMTO-ST Institute, Department of Applied Mechanics,\\ 
		CNRS/UFC/ENSMM/UTBM, University of Bourgogne Franche-Comté,\\
		\medskip
		24 Chemin de l'\'Epitaphe, 25000 Besançon, France. \\
		\textbf{Etienne Rivet $^3$} \\
		$^3$ Laboratoire de traitement des signaux 2\\
		École polytechnique fédérale de Lausanne, Lausanne, Switzerland
}


\title{Broadband nonreciprocal acoustic propagation using programmable boundary conditions: from analytical modelling to experimental implementation}

	\maketitle

	\begin{abstract} \noindent
		In this paper, we theoretically, numerically and experimentally demonstrate the acoustic isolator effect in a 1D waveguide with direction dependent controlled boundary conditions. A theoretical model is used to explain the principle of non reciprocal propagation in boundary controlled waveguides. Numerical simulations are carried out on a reduced model to show the non-reciprocity as well as the passivity of the system, through the computation of the scattering matrix and the power delivered by the system. Finally, an experimental implementation validate the potential of programmable boundary conditions for non reciprocal propagation.
	\end{abstract}

	\section{Introduction}
	
	In one-dimensional (guided) media, acoustic waves behave identically in both directions: the propagation of such waves has the property of reciprocity, owing to the symmetrical nature of physical laws under time and space reversal. Reciprocity is a fundamental property of wave equations ruling dynamic equilibrium of different physical phenomena as electromagnetism, mechanics or acoustics. Though uncommon in nature, direction-dependent propagation is sometimes a desirable feature: it provides isolation capabilities, which is of particular interest in all fields of physics. In electromagnetism, it has been shown that reciprocity can be broken in several ways: using biasing techniques like the Faraday effect \cite{aplet:1964}, nonlinear techniques \cite{krause:2008,poulton:2012,tocci:1995,gallo:2001}, or time-dependent modulation of physical properties \cite{lira:2012,yu:2009}.
	
	The acoustic isolator, also called acoustic diode, is a device where acoustic waves propagate in one direction, but blocks their transmission in the reverse direction, similarly to an optical isolator \cite{jalas:2013}. Although the term ``diode'' is often used to refer to this type of system, the isolator differs from the usual electrical diode. In the electrical-acoustical direct analogy where the electrical current $i$ is associated to the acoustic velocity $v$, the rectifier diode allows one half-period of harmonic acoustic wave to pass in one direction, while blocking the other half-period in the other direction, generating DC and second harmonic components. Note that such a diode has been patented long ago \cite{brevet_diode}.
	
	Breaking the reciprocity can be achieved through three known means: strongly nonlinear media with higher harmonic conversion \cite{liang:2009,liang:2010,boechler:2011,popa:2014}, static bias such as circulating fluid \cite{Fleury:2014, zangeneh2018doppler} (the acoustic equivalent of magnetic biasing in EM), and time-varying properties \cite{fleury:2015}. Other concepts relying solely on geometric features have been proposed, but taking into account all the inputs and outputs of the system shows that such concepts actually do not break reciprocity \cite{Maznev:2013}.
	
	On the one hand, most nonlinearity-based devices are usually limited by the necessary high input levels for the up-frequency conversion to be efficient, otherwise very poor transmission is obtained in the transmit direction. On the other hand, biased, resonant-type devices are inherently very narrow-band. The present paper aims at addressing these two points, using a concept of programmable boundary control.
	
	The outline of this contribution is as follows. First, from a general 3-dimensional model of guided waves in a duct, we formulate a one-dimensional theoretical model. Then, we numerically study a finite-length one-dimensional (1D) acoustic isolator based on this model, and characterize it in terms of its scattering matrix. Finally, we show experimental results obtained by applying a non-local boundary control on an acoustic waveguide, by means of digitally controlled electroacoustic devices.

	\section{Analytical modelling\label{sec:theory}}
	
	\subsection{Three-dimensional acoustic propagation in a waveguide}
\begin{figure}
		\centering
		\includegraphics[width=8.6cm]{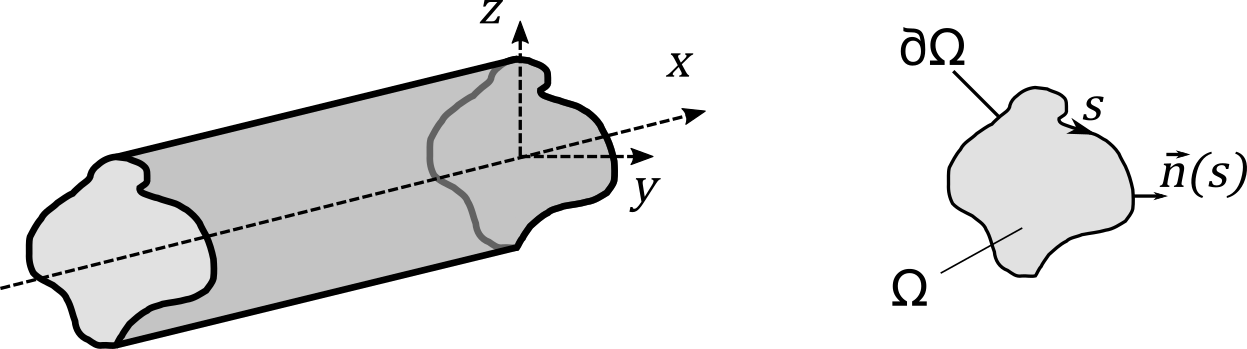}
		\caption{A cylindrical waveguide along coordinate $x$, with cross section of arbitrary shape $\Omega$. Left: overview of the waveguide. Right: detail of the cross-section and its contour $\partial\Omega$ parametrized by a curvilinear coordinate $s$. $\vec{n}$ is the local exterior normal at each point of the contour.}
		\label{fig:arbitrary_duct}
	\end{figure}

	We consider a cylindrical waveguide with a cross-section of arbitrary shape $\Omega\in\R^2$, and area $S$ (Fig. \ref{fig:arbitrary_duct}). The longitudinal coordinate along the waveguide is denoted $x$, transverse directions are denoted $(y,z)$. The closed contour around the cross-section is noted $\partial\Omega$.
	
	Inside such a waveguide, the linearized acoustic equations lead to the wave equation:
	$\forall t\in \R, \; \forall x\in\R$, and $(y,z)\in\Omega$:
	\begin{equation}
	\frac{1}{c_0^2}\partial_{tt} p(x,y,z,t) - \nabla^2 p(x,y,z,t) = 0 ,
	\end{equation}
	where $c_0=\sqrt{\chi/\rho_0}$ is the speed of sound in the host medium, with $\rho_0$ its mean density and $\chi$ its bulk modulus.
	The equation relating acoustic pressure and velocity is given by Eq. \ref{Euler_eq}:\\
	\begin{equation}\label{Euler_eq}
	\vec{\nabla} p(x,y,z,t) = -\rho_0 \partial_t \vec{v}(x,y,z,t).
	\end{equation}

	\subsection{Boundary control of wave propagation}
	
	Projecting Eq. \ref{Euler_eq} along the normal direction of the domain boundary $\partial\Omega$ expresses the boundary condition by a relation between the outgoing normal velocity $v_n$ and the acoustic pressure $p$. In case of a hard-walled waveguide, the normal velocity $v_n$ of the boundary is 0. In case of locally reacting walls characterized by a given acoustic impedance, the normal velocity $v_n$ is related to the local acoustic pressure $p$. This is typically the case of conventional absorbing materials, Helmholtz resonators, and other passive linings. In the case of an active lining \cite{karkar:2015a}, the normal velocity $v_n$ could be related to a more complex function of the acoustic pressure $p$ and its time or spatial derivatives. In the latter case, the walls could be modulated by a generalized impedance, which is a non-local operator.
	
	We now propose to use the following boundary condition on the walls of the waveguide, introduced by \cite{collet:2009}:
	\begin{equation}\label{eq:bndcontrol}
	\rho_0 \partial_t v_n = -\partial_n p = \frac{1}{c_a} \partial_t p - \partial_x p
	\end{equation}
	where $c_a$ is an advection celerity, which is used as a tunable parameter of the control. Although this partial derivative equation was meant as a boundary condition for the anomalous scattering of waves in interaction with an active surface \cite{collet:2009}, we use it here as a special boundary condition for a waveguide, where we expect waves to be under grazing incidence in the low frequency regime.

	\subsection{Reduction to 1D}
	To reduce this model to one dimension, we average the equations by integrating over the cross-section:\\
	$\forall t\in \R,  \quad x\in\R$, and $(y,z)\in\Omega$,
	\begin{equation}
	\frac{1}{S} \iint_\Omega \big[ \frac{1}{c_0^2} \partial_{tt} p(x,y,z,t) - (\partial_{xx} + \partial_{yy} + \partial_{zz}) p(x,y,z,t) \big] dydz = 0.
	\end{equation}
	
	Denoting the mean acoustic pressure over the section $\tilde{p}(x,t)$=$\frac{1}{S} \iint_\Omega p(x,y,z,t)dydz$, and integrating by parts the second term (or, equivalently, using the divergence theorem, or Stokes' theorem), the equation now reads:\\
	$\forall t\in \R,  \quad x\in\R$,
	\begin{equation}\label{eq:1Dreduced_wave_eq}
	\frac{1}{c_0^2} \partial_{tt} \tilde{p}(x,t) - \partial_{xx} \tilde{p}(x,t) = \frac{1}{S}\oint_{\partial\Omega} \partial_n p(x,s,t) ds
	\end{equation}
	where $s$ is a curvilinear coordinate along $\partial\Omega$ and $\partial_n$ is the gradient along the outward direction.
	
	Inserting the boundary control law, expressed in Eq. \ref{eq:bndcontrol}, in the right-hand side in Eq. \ref{eq:1Dreduced_wave_eq}, it reads:
	\begin{equation}
	\frac{1}{c_0^2} \partial_{tt} \tilde{p}(x,t) - \partial_{xx} \tilde{p}(x,t) = -\frac{1}{S}\oint_{\partial\Omega} \Biggr(\frac{1}{c_a} \partial_t p(x,s,t) - \partial_x p(x,s,t)\Biggr) ds
	\end{equation}
	
	Denoting the mean acoustic pressure field along the boundary contour $p_b(x,t)$=$\frac{1}{L_p} \oint_{\partial\Omega} p(x,s)ds$, where $L_p$ is the perimeter length of the cross-section, we finally get:
	\begin{equation}\label{eq:1Dreduced_wave_eq_withQ}
	\frac{1}{c_0^2} \partial_{tt} \tilde{p}(x,t) - \partial_{xx} \tilde{p}(x,t) = \partial_t Q_m(x,t)
	\end{equation}
	where the source term in the right-hand side is:
	\begin{equation}
	\partial_t Q_m = -\frac{1}{d} \Biggr( \frac{1}{c_a} \partial_t p_b (x,t) - \partial_x p_b(x,t) \Biggr).
	\label{src_term}
	\end{equation}
	with $1/d = L_p/S$ being now a second tuning parameter of the control, together with the advection celerity $c_a$.

	\subsection{Dispersion relation}\label{sec:dispersion}
	We now suppose monomodal acoustic propagation in this waveguide, and drop the tilda above $p$ and assimilate $p_b$ and $p$. Assuming a time-harmonic wave propagation with angular frequency $\w$ and wavenumber $k$, we now seek solutions of the form:
	\begin{equation}
	p(x,t) = p_0 e^{j(\w t-kx)}.
	\label{eq:wave}
	\end{equation}
	Inserting this ansatz in the propagation equation and in the special source distribution expressed in Eqs. \ref{eq:1Dreduced_wave_eq_withQ} and \ref{src_term} respectively, comes:
	\begin{equation}
	(j\w/c_0)^2 p(x,t) - (-jk)^2 p(x,t) = -\frac{1}{d}\Biggr((j\w/c_a) p(x,t) -(-jk) p(x,t)\Biggr)
	\end{equation}
	which, being true for all $t\in\R$ and all $x\in[0,L]$, implies the following relationship:
	\begin{equation}
	(j\w/c_0)^2 - (jk)^2 = -\frac{1}{d}(j\w/c_a+jk).
	\label{eq:dispersion}
	\end{equation}
	
	In the special case where $c_a=c_0=c$ (that is the advection celerity is in tune with the sound speed), we get:
	\begin{equation}\label{eq:dispersion_c0}
	j\w/c(j\w/c+1/d) = jk(jk-1/d)
	\end{equation}
	which has two obvious solutions: either $k^{(-)}=-\w/c$, or $k^{(+)}=\w/c-j/d$.
	
	The first solution corresponds to waves with purely real and negative wave numbers. The group velocity is negative, and equal to the phase velocity: this type of wave is propagating in the (-) direction ("backward waves"). Given the purely real wavenumber, they pass through the isolator without attenuation.
	
	Inserting the second solution in the ansatz expressed in Eq. \ref{eq:wave}, we write:
	\begin{equation}
	p(x,t) = p_0 e^{j\w(t-x/c)}e^{-x/d}.
	\end{equation}
	This second solution corresponds to waves with $\Re(k)>0$ that propagate in the (+) direction ("forward waves"), given their positive group velocity, but with an exponential attenuation. It is equivalent to an evanescent guided mode
	, just like a waveguide higher mode under its cutoff frequency.
	
	We can now physically interpret the meaning of $d$: it is the typical attenuation length. If it is short enough, a good acoustic isolator can be designed: sound waves can only pass in one direction through the isolator.
	

	\subsection{Acoustic power delivered by the source}\label{sec:acoustic power analytical}
	Given the source term $Q_m$, the power injected by the isolator into the acoustic domain is obtained as follows:
	\begin{equation}
	\mathcal{P}(x,\w) = \frac{1}{2}\Re \big[Q_m(x,\w) p^*(x,\w)/\rho_0 \big]
	\label{eq:power_product}
	\end{equation}
	where $\mathcal{P}$ stands for the active acoustic power density, or acoustic intensity per unit length.
	
	Using the ansatz \ref{eq:wave} and the definition \ref{src_term} of the source term, comes:
	\begin{equation}
	\mathcal{P}(x,\w) = -\frac{1}{2d} \frac{|p(x,\w)|^2}{\rho_0 c} \Re \big[\frac{j\w/c + jk}{j\w/c}\big]
	\label{eq:power}
	\end{equation}
	
	Clearly, for backward waves, where $k$=$-\w/c$, the power density of the source vanishes because of the numerator of the rightmost fraction. The source is inactive.
	
	For forward waves, inserting the dispersion relation of Eq. \ref{eq:dispersion_c0} into Eq. \ref{eq:power}, the power density is:
	\begin{equation}
	\mathcal{P}(x,\w) = -\frac{1}{d} \frac{p_0^2}{\rho_0 c} e^{-2x/d}
	\end{equation}
	This value is always negative, for every $x$ and every $\w$, so the source is always absorbing acoustic power: the system is acoustically passive for these waves.
	
	However, due to internal reflections, when a forward incident wave enters the isolator, the pressure field in the isolator actually consists of a superposition of forward and backward waves. The forward wave cannot occur separately from the backward wave. Even though backward waves alone do not contribute to the source term, it does have an incidence on the power balance in case of a superposition of both forward and backward waves, as it contributes to the pressure term of the product in Eq. \ref{eq:power_product}. It is not obvious, in that case, to prove that the source is passive, so this point will be investigated in the next section, by means of numerical simulations.

	\section{1D Numerical Simulation}\label{sec:numerical}

	\subsection{Modelling}
	\begin{figure}
		\centering \label{fig:1Dwaveguide}
		\includegraphics[width=8.6cm]{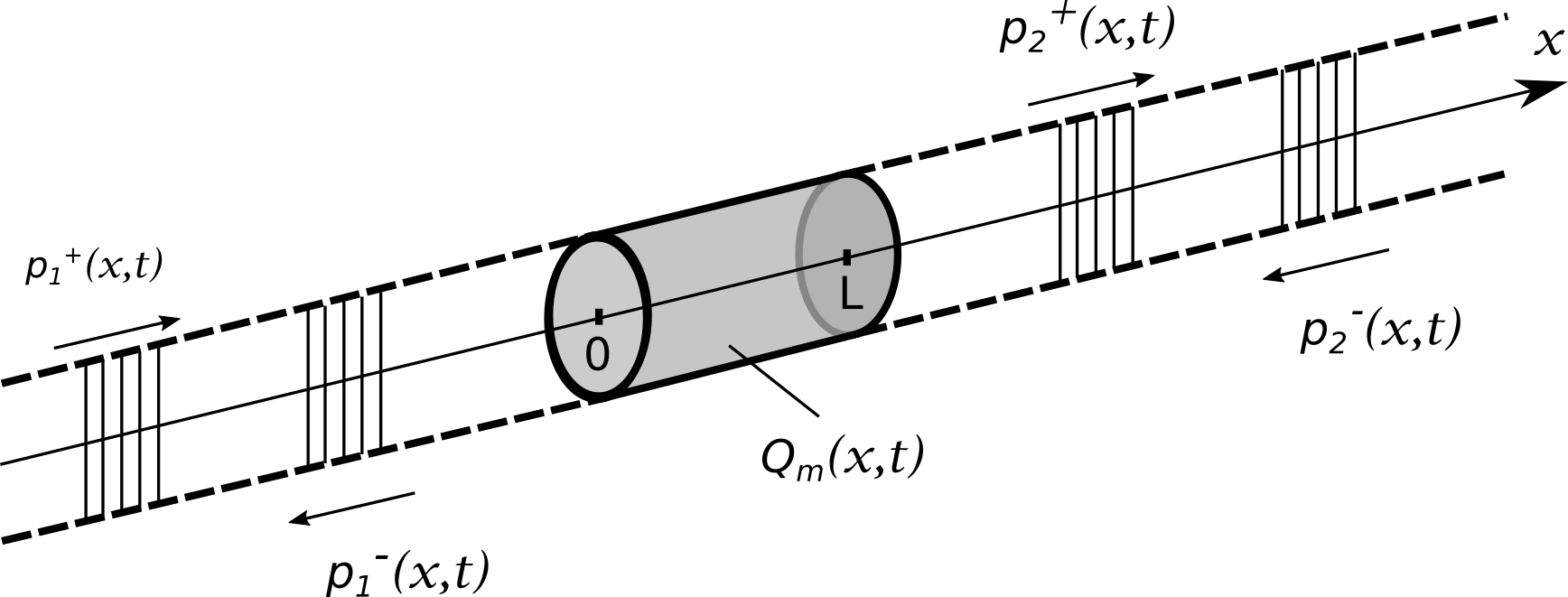}
		\caption{A finite-sized acoustic isolator connected to semi-infinite waveguides at both ends. Ingoing and outgoing waves are shown on both sides.}
		\label{fig:schema}
	\end{figure}
	
	We simulated the theoretical model proposed in Section \ref{sec:theory} using a finite element software in the frequency domain, and meshing the isolator uniformly with a step $h$=3.46 mm (sufficient to resolve acoustic waves up to wavelength $\l$=$10h$, or $f$=10 kHz). The length $L$ of the isolator was equal to 0.33 m in accordance with the experimental implementation presented in Section \ref{sec:experimental}. The medium was dry air at standard temperature and pressure. The density used for the medium was $\rho_0$=1.18 kg/m$^3$ and speed of sound was $c_0$=346 m/s.

	We applied a plane wave radiation boundary condition at the left-most and right-most ends of the system to realize anechoic terminations, and we set $Q_m$ on the isolator, as defined in Eq. \ref{src_term}. We have also applied a 1 Pa RMS incident plane wave (94 dB ref. 20 $\mu$Pa) at the waveguide input. The total acoustic field has been resolved in the whole system, for frequencies ranging from 10 Hz to 10 kHz.

	
	\subsection{Scattering-matrix}
	We now aim at characterizing this isolator in terms of its scattering matrix $S$. In the present case, $S$ is a two by two matrix, linking outgoing waves to ingoing waves in the left (subscript $1$) and right (subscript $2$) regions outside the device (see Fig. \ref{fig:schema}).
	
	\begin{equation}
	\begin{bmatrix}
	p_1^-\\p_2^+
	\end{bmatrix}
	= \begin{bmatrix} S_{11} & S_{12} \\ S_{21} & S_{22} \end{bmatrix}
	\begin{bmatrix}
	p_1^+\\
	p_2^-
	\end{bmatrix}
	\end{equation}
	
\begin{figure*}
		\centering
		\includegraphics[width=.45\textwidth]{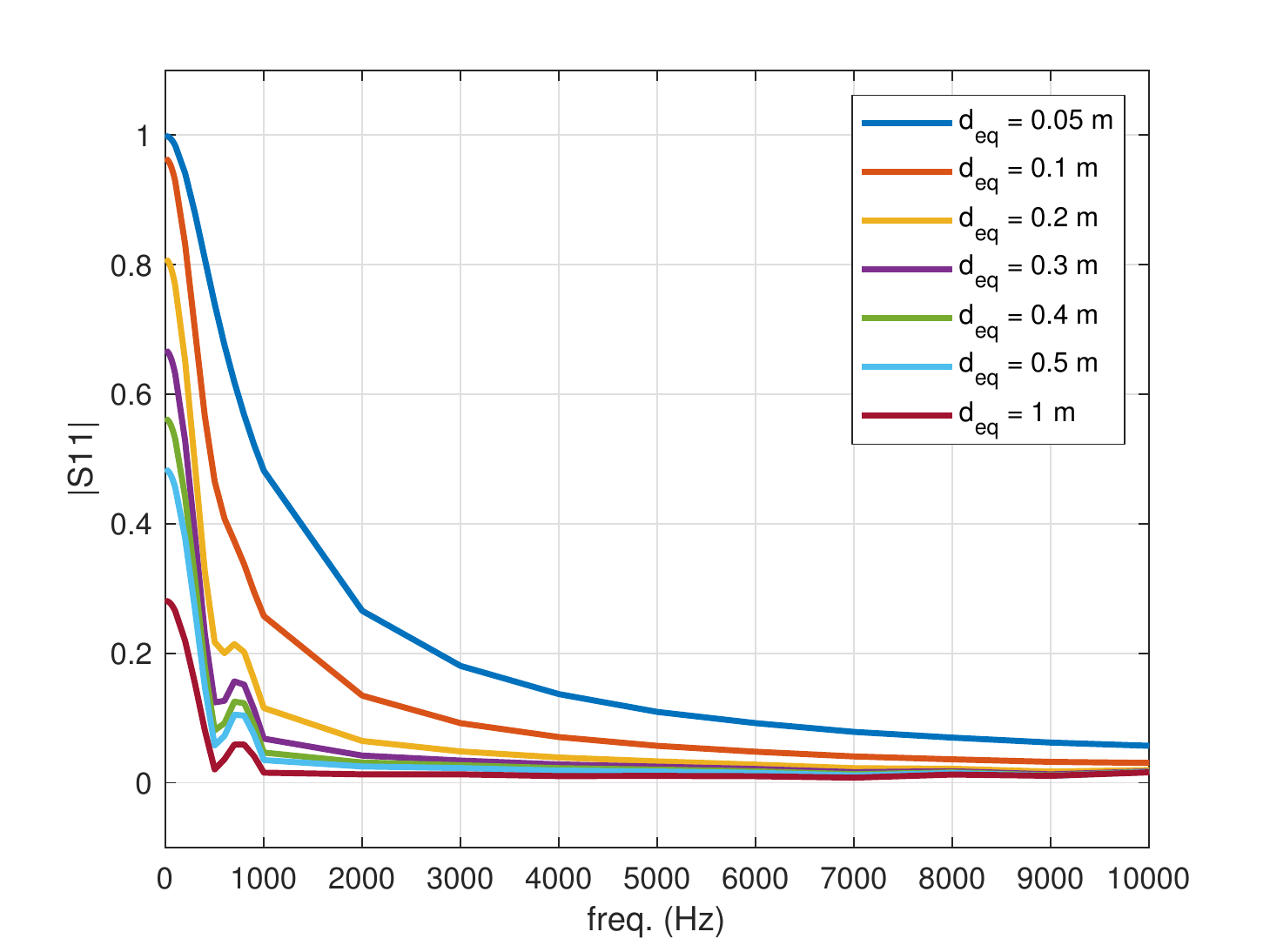}\includegraphics[width=.45\textwidth]{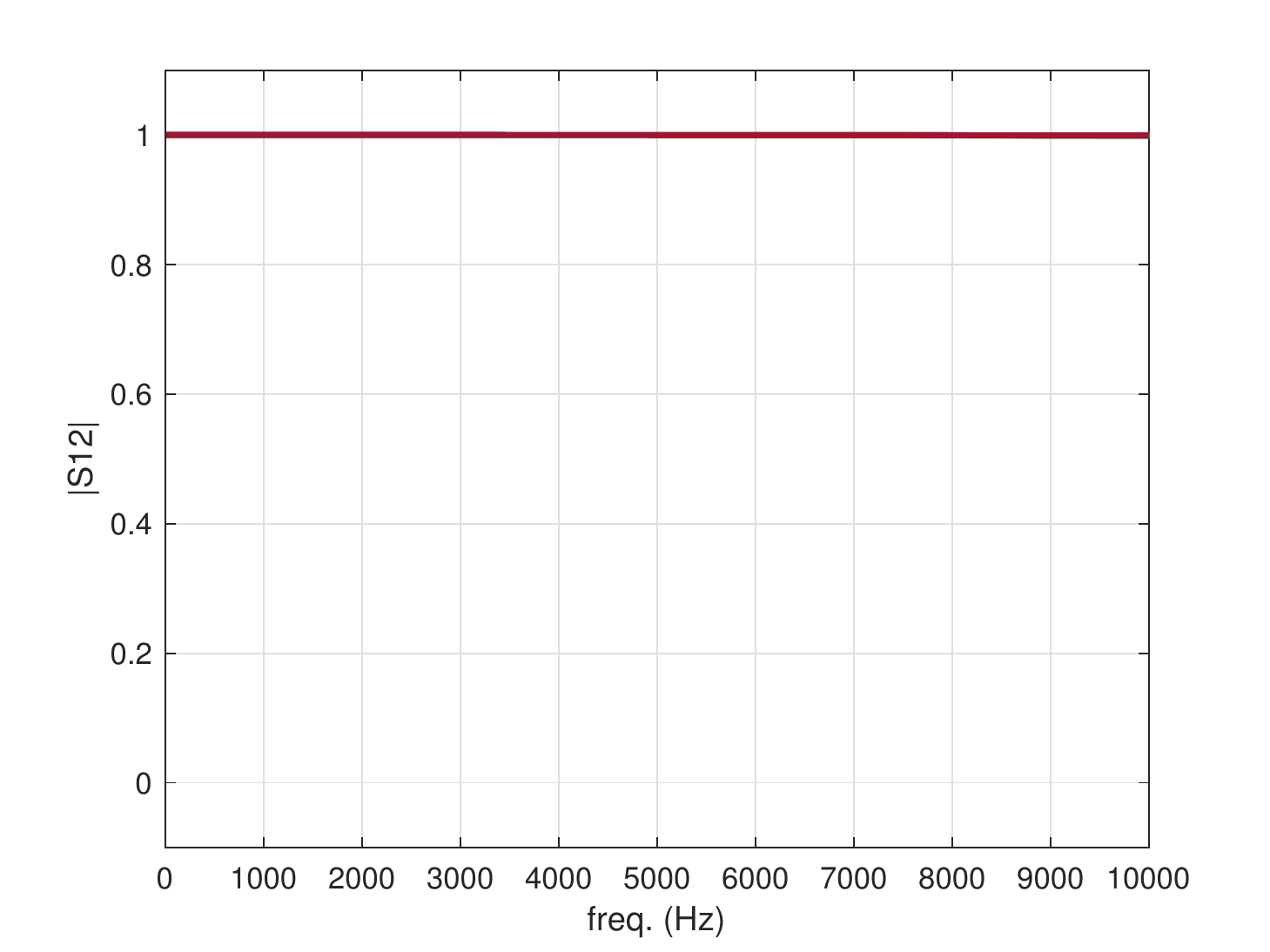}\\
		\includegraphics[width=.45\textwidth]{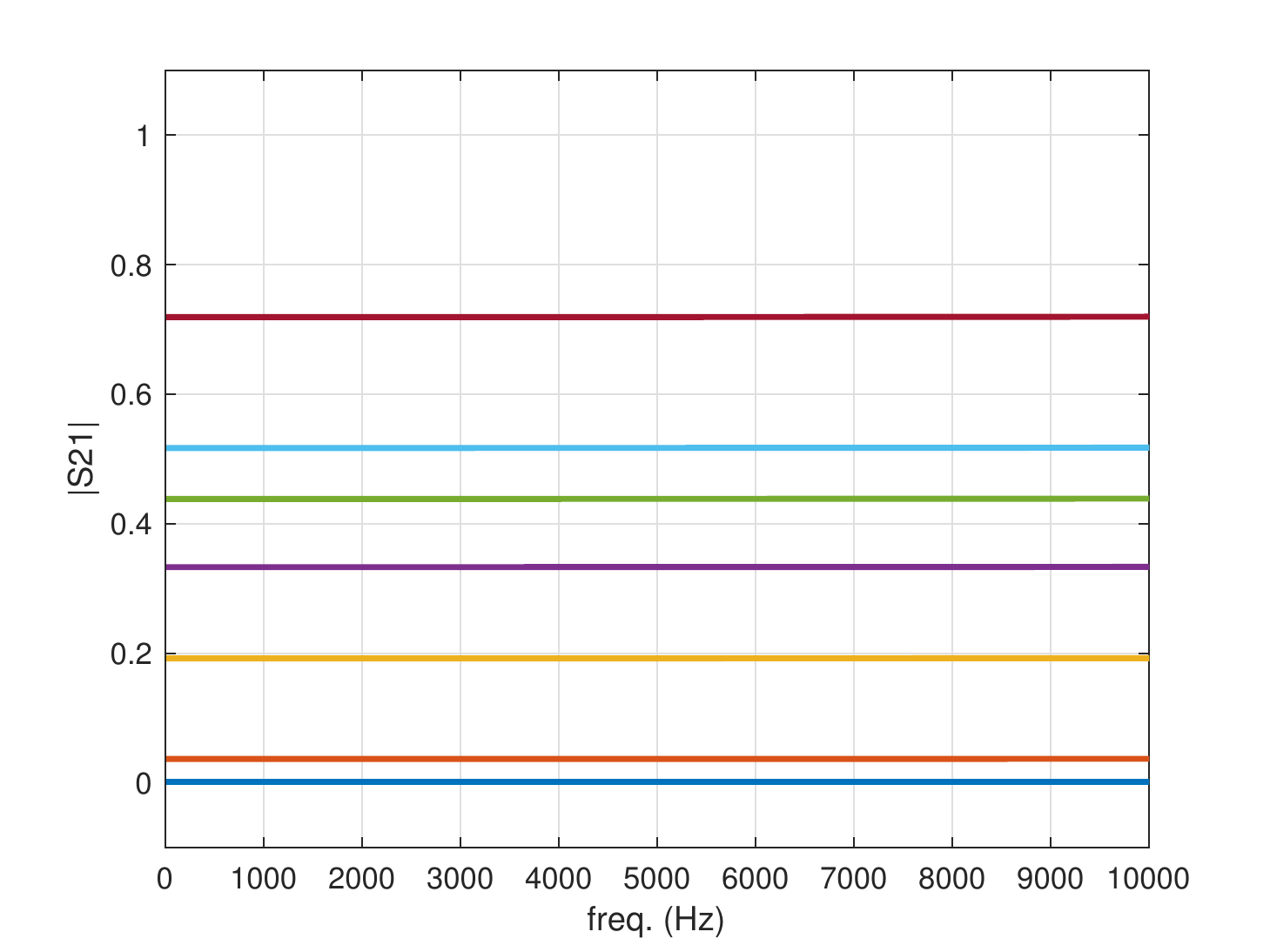}\includegraphics[width=.45\textwidth]{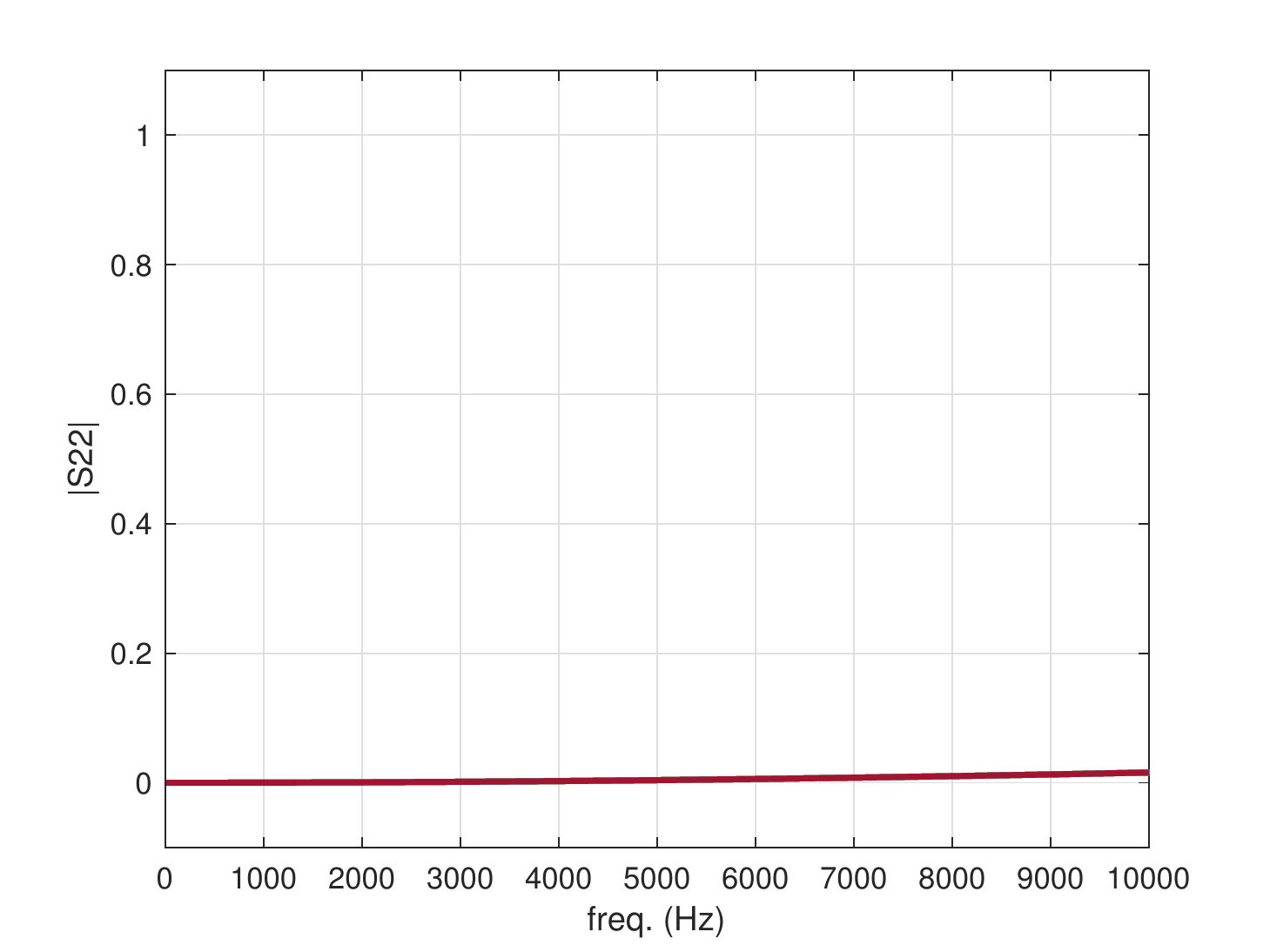}
		\caption{Scattering matrix coefficients magnitude (dimensionless) of the proposed acoustic isolator as a function of frequency (in Hz), for different parameter values $d$. Top-left: $|S_{11}|$, top-right: $|S_{12}|$, bottom-left: $|S_{21}|$, and bottom-right: $|S_{22}|$.} 
		\label{fig:Smatrix}
		
	\end{figure*}

	The scattering matrix coefficients are presented in Fig. \ref{fig:Smatrix} by representing the magnitude (in dB) of each coefficient as a function of the frequency, for several values of $d$. As expected, $S_{12}$ is very close to 1, and $S_{22}$ to 0, whatever the value of $d$ and the frequency, which confirms the perfect transmission for backward waves. $S_{21}$ follows exactly the expected value discussed in Section \ref{sec:dispersion}, i.e. exp($-L/d$), whatever the frequency. And the reflection coefficient $S_{11}$ for forward waves decreases with the frequency as well as with the parameter $d$. This is to be expected if we consider that $d = S/L_p$ is actually a characteristic dimension of the cross section of the duct: the larger the cross section dimension, the less efficient any liner is in terms of transmission attenuation.
	
	%
	%
	%
	%
	
\begin{figure}
		\centering
		\includegraphics[width=8.6cm]{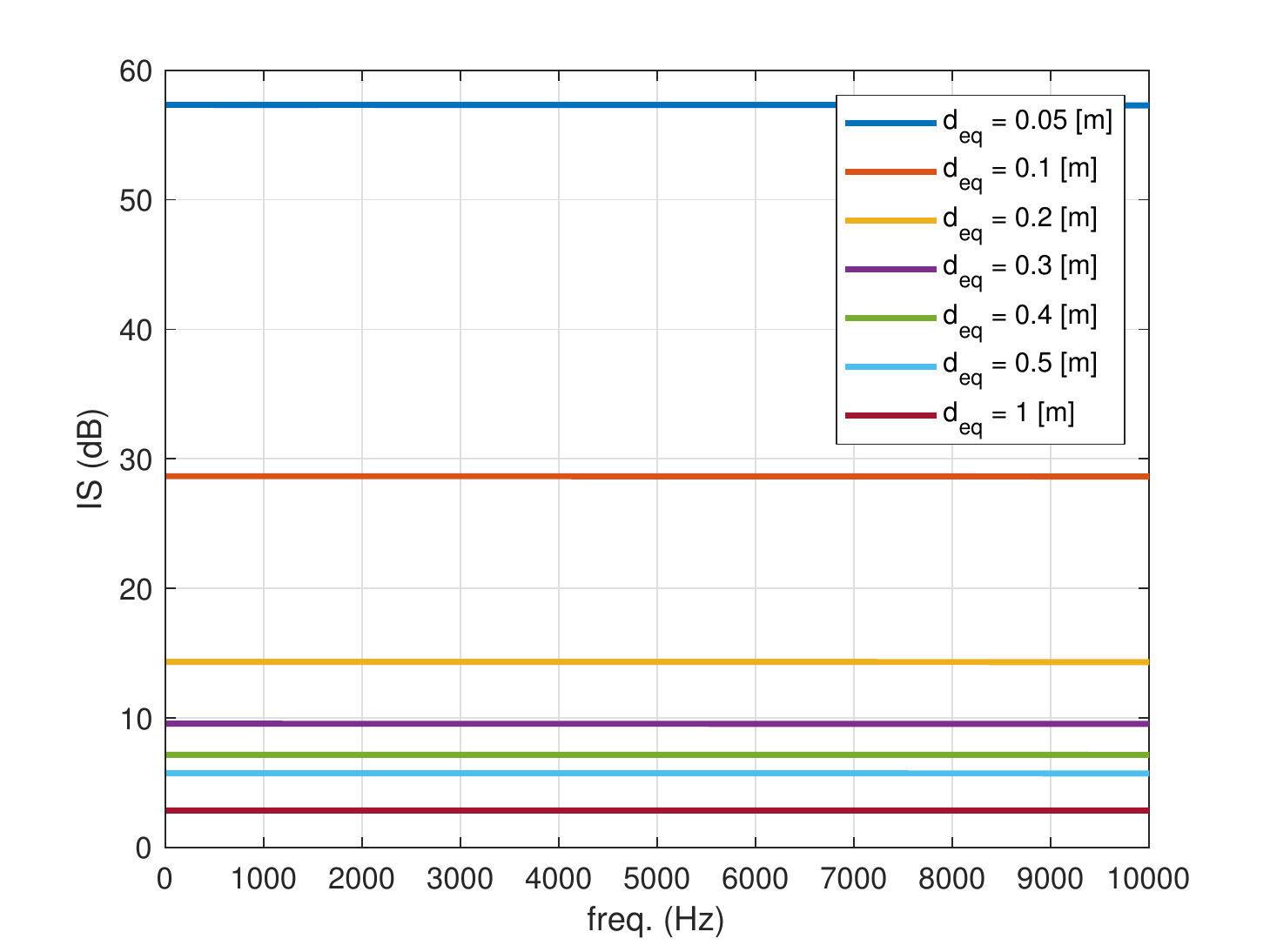}
		\caption{Isolation (in dB) as a function of frequency, for different values of the parameter $d$.}
		\label{fig:IS_numerical}
	\end{figure}

	The calculation of the Isolation Index, $IS = 20 \log_{10}(|S_{12}/S_{21}|)$, is straightforward and its plot is in Fig. \ref{fig:IS_numerical}. For every value of $d$, the isolation is constant over the whole frequency range and proportional to $L/d$. Hence, the required isolation for a given application can be specified and tuned accordingly by working either on the length $L$ of the isolator, or on the parameter $d$.  
	
	\subsection{Acoustic power delivered by the source}\label{sec:acoustic power numerical}
\begin{figure}
		\centering
		\includegraphics[width=8.6cm]{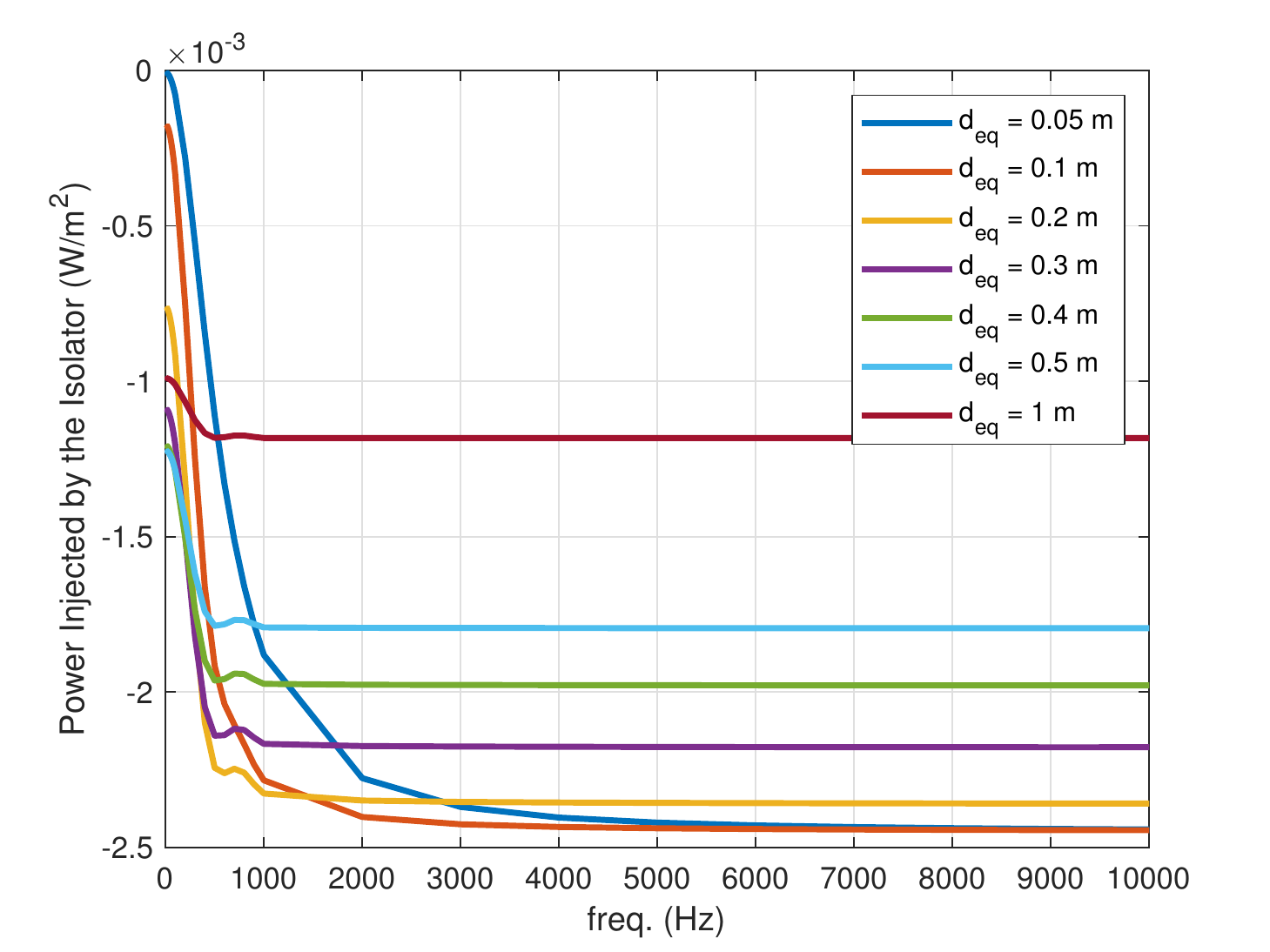}
		\caption{Acoustic power injected by the isolator into the acoustic domain, as a function of frequency, for different values of the parameter $d$, in case of an incident wave of $1$Pa RMS in the positive $x$ direction.}
		\label{fig:1D-Psource-freq}
	\end{figure}

	The power computed in post-processing, that is illustrated in Fig. \ref{fig:1D-Psource-freq} for incident wave in the positive $x$ direction, shows that the distributed source never brings energy to the system, as expected. The total power injected by the isolator is defined as $P_{isolator}(\w) = \int_{x=0}^{L}P(x,\w)dx$, where $P(x,\w)$ is defined in Eq. \ref{eq:power_product}
	
	In Fig. \ref{fig:1D-Psource-freq} we observe that the system is acoustically passive whatever the value of the parameter $d$. For incident waves from the opposite side of the isolator (propagating in the negative x direction), the total power injected by the isolator is identically equal to 0 (not shown here).
	
	\subsection{Pressure level profile}
	\begin{figure}
		\centering
		\includegraphics[width=8.6cm]{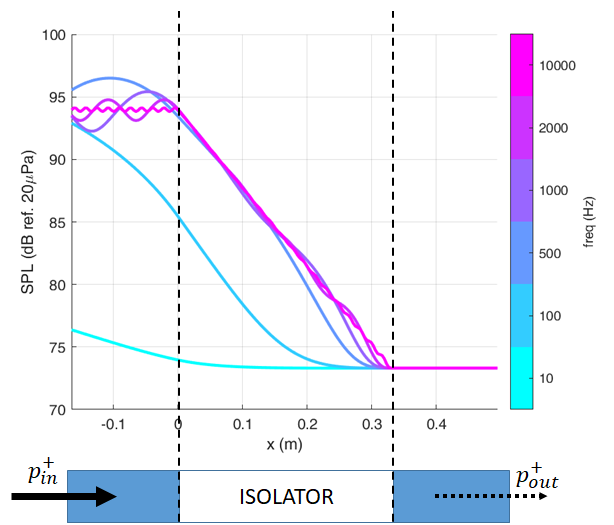}
		\caption{Sound pressure level (in dB ref.20$\mu$ Pa) along the system for different frequencies, in case of an incident wave of $1$ Pa RMS in the positive $x$ direction. Parameter $d=0.14$ m. The isolator is located between $x=0$ and $x=0.33$ m}
		\label{fig:1D-Lp}
	\end{figure}
	Figure \ref{fig:1D-Lp} shows the sound pressure level $SPL$ (in dB) along the system axis, for several frequencies, and for an incident wave of $1$Pa RMS in the positive x-direction. The parameter $d$ is set to 0.14 m. This value of the parameter $d$ corresponds to the experimental implementation of Section \ref{sec:experimental}. The linear decrease of $SPL$ with $x$ confirms the exponential decay of the incident waves in the isolator. A reflected wave is created resulting in interference patterns at the waveguide input, and frequency-dependent amplitudes for the sound pressure level.
	
	The equivalent figure for retrograde incident waves (not shown here), is perfectly flat and null: the waves propagate as in a rigid waveguide.

	\section{Experimental implementation}\label{sec:experimental}
	
	\subsection{Test-bench}\label{sec:experimental setup}
	
\begin{figure}
		\centering
		\includegraphics[width=.9\textwidth]{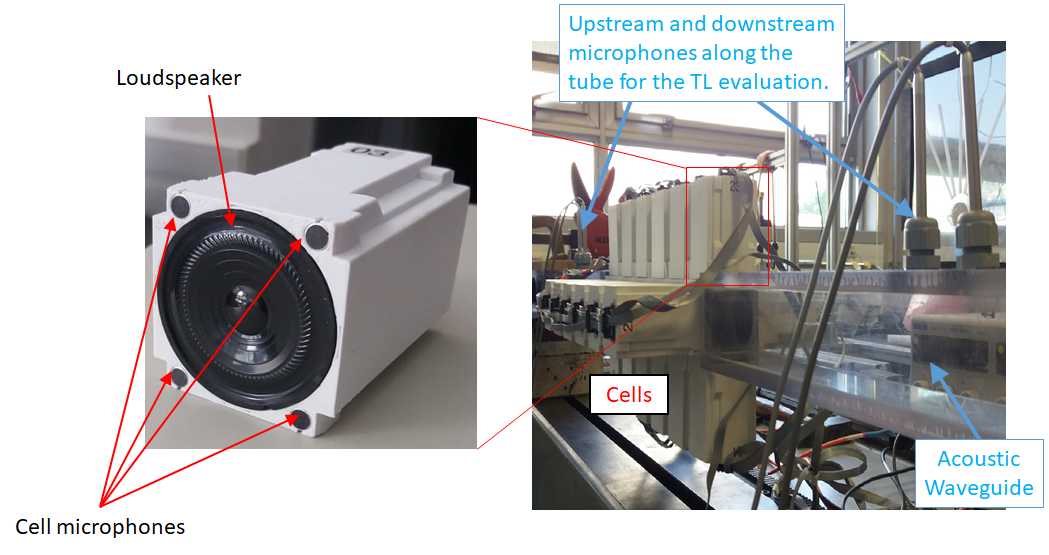}\\
		\caption{Left: Unit cell. Right: Prespective of the lined waveguide, with upstream and downstream microphones for the TL evaluation through the transfer matrix method.}
		\label{fig:ExpSetup}
	\end{figure}

	The non-local boundary control isolation has been proved thanks to an analytical and numerical reduced model. We now present the experimental results of an application of this boundary control strategy on a test-bench waveguide of square cross section of 5.5cm side (see Fig. \ref{fig:ExpSetup} on the right), lined with actively controlled devices which were capable of reproducing the boundary control law expressed in Eq. \ref{eq:bndcontrol}. Evidently, the boundary control law has been discretized in order to be applied through units of electrodynamic devices. Each unit cell consisted of a loudspeaker (actuator) and four microphones (sensors) placed at the cell corners, see Fig. \ref{fig:ExpSetup} on the left. Its discretization and the algorithm for the control implementation are reported in Appendix \ref{app:exp details}.\\
	Six cells have been applied on each side of the duct, amounting to 24 cells in total. The size of each cell was approximately equal to the lateral side of the duct section and the lined portion of duct was $L = 33$ cm.\\
	Both ends of the tube were filled with 45cm of foam in order to reproduce anechoic conditions at the input and output of the waveguide. The acoustic source has been placed flush with the duct surface just ahead of the foam termination, sufficiently far from the lined portion of duct in order for plane waves to be fully developed before the isolator. The transmission loss measurements have been performed according to the ASTM E2611-09 standard (single-load transfer matrix method) and the Insertion Loss has been retrieved from the hard-wall duct case. 
	
	\subsection{Results}\label{sec:experimental results}
	
\begin{figure}
		\centering
		\includegraphics[width=.45\textwidth]{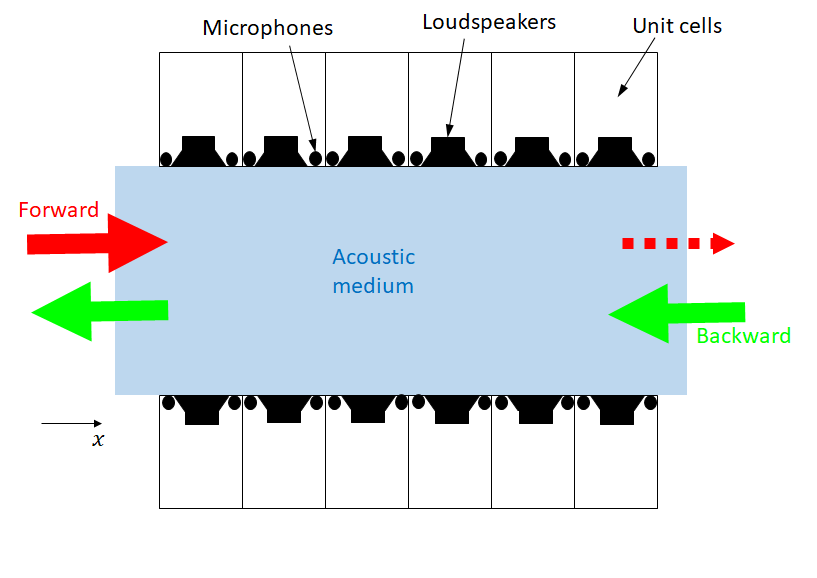}\includegraphics[width=.45\textwidth]{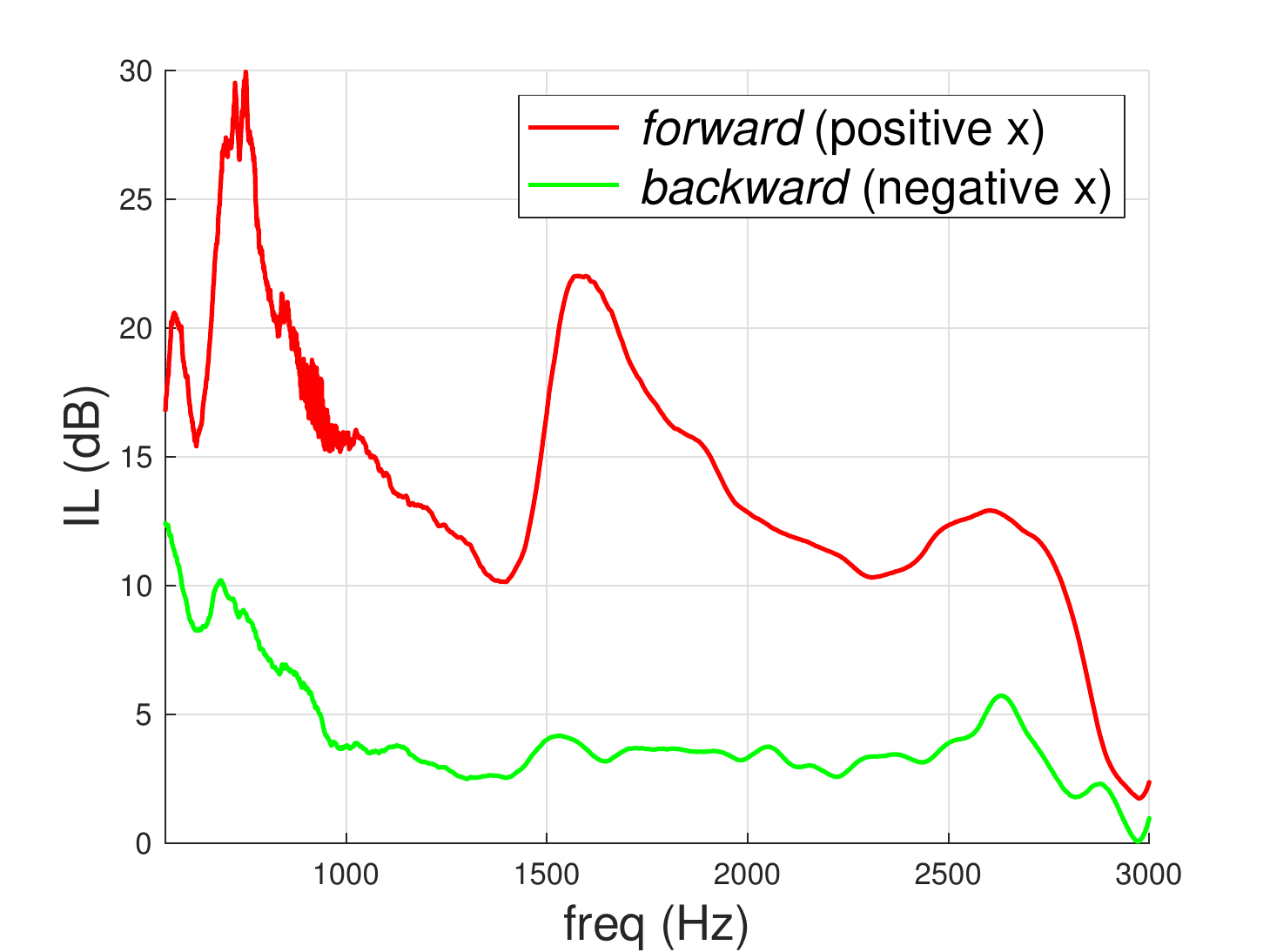}
		\caption{
			Left: 2D sketch of the lined waveguide. Right: Insertion Loss measurements relative to incident waves directed towards the positive x (\emph{forward}), and towards the negative x (\emph{backward}).}
		\label{fig:sketch and IL}
	\end{figure}

	The ILs are plotted in Fig. \ref{fig:sketch and IL} (on the right), for incident waves coming from both directions.\\
	The synthesized algorithm for the implementation of the control law was able to invert the dynamics of the loudspeaker after its first resonance (see Appendix \ref{app:exp details}) which is between 500 and 550 Hz for all loudspeakers, that is why the plot is limited to frequencies beyond 550 Hz. The higher frequency limit (3 kHz) corresponds to the cut-on of the first higher-order duct mode, beyond which the plane wave assumption for the TL evaluation is not valid anymore.
	
	\begin{figure}
		\centering
		\includegraphics[width=8.6cm]{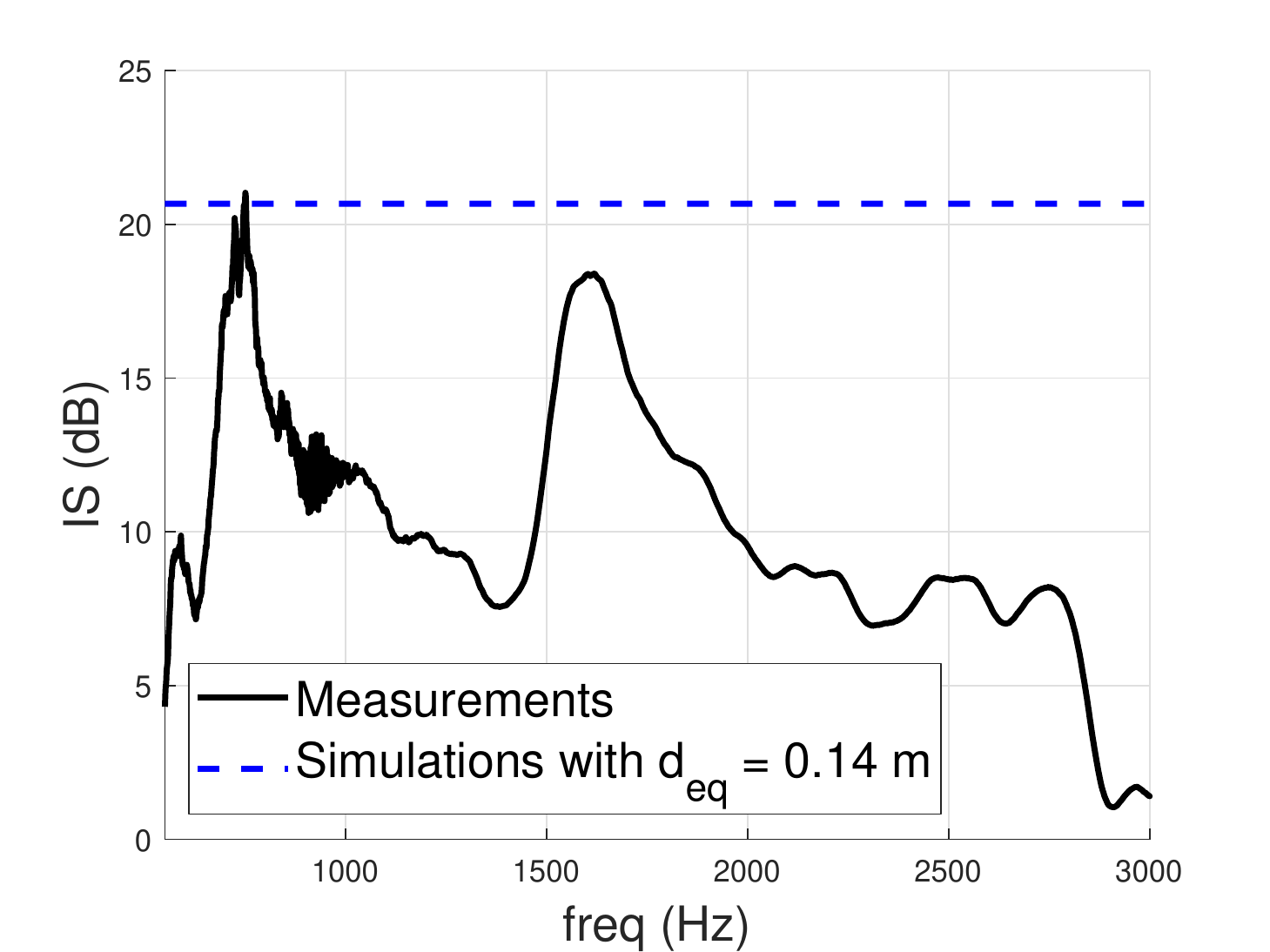}\\
		\caption{Isolation index: measurements compared to 1D reduced model simulation with $d$ equivalent to the experimental application.}
		\label{fig:IS}
	\end{figure}
	
	The few dBs of attenuation for the \emph{backward} direction of propagation are mainly due to two causes. First, the celerity coefficient $c_a$ adopted in the control was not taken exactly equal to the speed of sound $c_0$, but slightly lower. The choice of $c_a<c_0$ was made in order to avoid sound radiation from the liner into the acoustic domain, which happens for $c_a>c_0$ in an analogous way to the coincidence frequency phenomenon of fluid-structure interaction. A study of the effect of the variation of $c_a$ on stability is reported in \cite{debono2018}. Second, we note that the dynamic model of the loudspeaker used in the control, assumes a simple inertial behaviour (see Appendix \ref{app:exp details}), which did not allow the control to annihilate the speaker membrane flexibility at all frequencies.\\
	The Isolation Index IS retrieved from the ILs along the two opposite directions of incident waves, is reported in Fig. \ref{fig:IS}. It is compared to the IS predicted by the 1D reduction model outlined in Sections \ref{sec:theory} and \ref{sec:numerical}, with a parameter $d = d_{eq} = 0.14$ m. The value $d_{eq}$ relative to an equivalent 1D reduced model of our actual experimental application, is calculated in Appendix \ref{app:equivalent gain}.\\
	The IL relative to the \emph{forward} direction of propagation and the IS vary significantly with frequency. This is due to the modal behaviour of the speaker membrane and the limits of the synthesized control to invert all the loudspeaker dynamics. As a consequence, the acoustic isolation is achieved with a variable amplitude: from maxima of 20 dB to minima of 8 dB, from 550 Hz up to 2800 Hz. The loss of efficiency above 2800 Hz is likely to be mainly caused by the spacing between the microphones, which is close to half the acoustic wavelength, thus causing aliasing effect in the pressure evaluation.\\
	It is interesting to notice the coincidence between the 1D numerical simulations and the experimental results at about 750 Hz. It is the frequency around which we have maximum efficiency of the control algorithm, i.e. the actuator inertial model we used in the control transfer functions (Appendix \ref{app:exp details}), match the actual mechanical dynamics of the loudspeaker, so that we reach the ideal value of isolation. The second peak at about 1.6 kHz instead corresponds to the second mode of the loudspeaker membrane which apparently results to be highly enhanced by our control.\\
	The maximum achievable amplitude of isolation depends upon the only parameter $d_{eq}$, which takes into account the size of the waveguide cross section, the percentage of surface lined by the discrete actuators, and the control algorithm gain $g$ (see Eq. \ref{eq:d_eq}), the last one being limited by delay-related passivity issues (see Appendix \ref{app:passivity_control}).
	
	
	\section{Discussion}
	
	Boradband acoustic isolation is here achieved through a new strategy, based upon a controlled non-locally reacting liner, which hardly identifies with any of the non-reciprocal acoustic devices reported in the review \cite{fleury2015nonreciprocal}.
	In an attempt to fit our technique into the categories of \cite{fleury2015nonreciprocal}, we may see it as a \emph{boundary} spatio-temporal modulated device, which determines a bias acoustic field \emph{in response to} an external excitation. As a proper (non-locally) \emph{reacting} liner, there is no bias field without external excitation. This last feature is very important, as it differentiates our method from flow-biased devices \cite{fleury:2015}, and makes it a good candidate for achieving non-reciprocal acoustic propagation in waveguides which must be travelled by air (as in ventilation systems, intakes and so on), or by any other noise-producing medium. The experimental implementation of such a biased boundary has been realized through a loudspeaker control, similarly to the active Acoustic Metamaterial in \cite{boulandet2018duct}. But, for the first time, the concept of reconfigurable metasurfaces \cite{zangeneh2019active} is here confronting the new challenge of non-reciprocal progation. Its broadband achievement opens up new horizons to sound wave-steering through boundary control.\\ 
	Its limitations are inherent to the active control chain itself. The loudspeaker model adopted in our first experimental implementation, along with the inevitable delay introduced by the digital control, constrain the performance of the device. These practical issues are being investigated in order to synthesize more robust algorithms leading to a larger bandwidth (in the lower frequencies) and higher isolation levels. The minimum frequency of efficient isolation, here related to the resonance frequency of the loudspeaker, can indeed be significantly lowered by either reducing the actuator resonance frequency itself through electrical shunting techniques \cite{lissek2012optimization}, or by a special filtering strategy \cite{boulandet2018duct}. On the other hand, the higher frequency limit of efficiency is imposed by the dimensions of the cells. For this reason new generation of smaller cells are being produced.\\
	Nonetheless, this first and yet non-optimized prototype of non-local boundary control proves the non-reciprocity with isolation levels ranging from minima of 8 dB up to maxima of 18 and 20 dB, from the subwavelength regime (at 600 Hz, the size of our cell is less than $\lambda/10$), up to 2800 Hz (therefore for more than 2 octave bands).

	\section{Conclusions}
	
	In this contribution we analitically and numerically demonstrated the acoustic isolation achieved by a non-local boundary control strategy. While providing good level of isolation, the particularity of the proposed approach is its inherent broadband nature, in comparison to usual resonance based metamaterials and bias techniques. The first experimental results reported here certifies the broadband nature of this approach to non-reciprocal acoustics. Future work will focuse on the optimization and robustness of the electro-mechano-acoustical control chain, to further enlarge the bandwidth and increase the level of acoustic isolation.

	\appendix
	\section{Boundary Control implementation through electroacoustic cells}\label{app:exp details}
	
	In order for the boundary control to be implementated through finite electroacoustic cells, the space differential operator appearing in the non-local boundary control of Eq. \ref{eq:bndcontrol} has been discretized using a first order Euler approximation, as in Eq. \ref{eq:bndcontrol_discretized}.
	
	\begin{equation}\label{eq:bndcontrol_discretized}
	\rho_0 \dot{v}_n^{(i)} = \frac{1}{c_a} \dot{p}^{(i)} - \frac{p_{down}^{(i)} - {p}_{up}^{(i)}}{dx}
	\end{equation}
	where $\dot{p}^{(i)}$ is the pressure time derivative averaged on the four microphones of the $i^{th}$ cell; $p_{down}^{(i)}$ and $p_{up}^{(i)}$ are the pressures averaged over the two microphones respectively downstream and upstream the $i^{th}$ loudspeaker; $dx\simeq 0.04$m is the spacing between the upstream and downstream microphones in the cells, which gives the upper limit of validity of the finite difference approximation.\\
	The electroacoustic cells showed in Fig. \ref{fig:ExpSetup} on the left, reproduce the boundary condition of Eq. \ref{eq:bndcontrol_discretized}, thanks to a programmable digital signal processor. The inputs of the controller were the measured averaged pressure $p$ and its gradient $\frac{dp}{dx}=\frac{p_{down}^{(i)} - {p}_{up}^{(i)}}{dx}$, and the output was an electrical current $i$ in the loudspeaker coil. The transfer functions between the inputs and output, were able to invert the loudspeaker proper dynamics, and reproduce the boundary law of Eq. \ref{eq:bndcontrol_discretized}. The loudspeaker mechanical dynamics in the frequency domain is described in Eq. \ref{eq:loudspeaker mechanical dynamics}, according to the single-degree-of-freedom approximation valid around the "piston-mode" resonance frequency \cite{beranek2012acoustics}.\\
	
	\begin{equation}\label{eq:loudspeaker mechanical dynamics}
	Z_m(\w) v_n(\w) = S_d \; p(\w) - Bl \; i(\w)
	\end{equation}
	where $v_n(\w)$, $p(\w)$ and $i(\w)$ are respectively the inward normal velocity at the speaker diaphragm, the local pressure and the electrical current in the loudspeaker coil, as functions of the pulsation $\w$; $Z_m$ is the mechanical impedance of the loudspeaker, $S_d$ is the effective piston-area and $Bl$ is the force factor. In turn, $Z_m = j\w M_m + R_m + \frac{1}{j\w C_m}$, where $M_m$, $R_m$, and $C_m$ are the mechanical mass, resistance and equivalent or total compliance. These are called Thiele-Small parameters of the loudspeaker.\\
	The target boundary law of Eq. \ref{eq:bndcontrol_discretized} is expressed in the frequency domain in Eq. \ref{eq:bndcontrol_freq}.
	
	\begin{equation}\label{eq:bndcontrol_freq}
	\rho_0 j\w v_n(\w) = \frac{1}{c_a} j\w p(\w) - \frac{dp}{dx}(\w) .
	\end{equation}
	
	In order to reproduce the control law in Eq. \ref{eq:bndcontrol_freq} on the loudspeakers diaphragms, the digital control must be able to synthsize the transfer functions reported in Eqs. \ref{eq:Hloc} and \ref{eq:Hdist} between the output $i(\w)$ and the inputs $p(\w)$ and $dp/dx(\w)$ respectively (see \cite{rivet2016room} for further insights on the $H_{loc}$ transfer function).
	
	\begin{equation}\label{eq:Hloc}
	\frac{i(\w)}{p(\w)} = H_{loc}(\w) = \frac{S_d}{Bl} - g \; BP(\w) \frac{1}{Bl} \frac{Z_m(\w)}{\rho_0 c_a}  
	\end{equation}
	
	\begin{equation}\label{eq:Hdist}
	\frac{i(\w)}{\frac{dp}{dx}(\w)} = H_{dist}(\w) = g \; BP(\w)\frac{Z_m(\w)}{Bl}\frac{1}{\rho_0j\w}
	\end{equation}
	where $BP(\w)$ is a band-pass filter properly designed in order to avoid current saturation (at low and high frequencies) as well as to exclude the frequency regions where the loudspeaker dynamics is not accurately inverted; $g$ is a constant gain.\\
	In the definition of the control transfer functions $H_{loc}$ and $H_{dist}$, all the Thiele-Small parameters illustrated above must be known, and their estimation and corresponding incertitude highly affect the performances of such a control. In the first experimental implementation reported in this paper, we adopted the above transfer functions but with a \emph{mass-like} loudspeaker mechanical impedance model, i.e. $Z_m = j\w M_m$. This assumption is acceptable for frequencies sufficiently beyond the loudspeaker resonance, and it follows up the former experimental implementation of the non-local boundary control of \cite{collet:2009}, even though in a different control strategy. While relieving us from the model incertitude issues at the resonance, and enlarging the efficiency of the non-local boundary control to higher frequencies, this impedance model assumption does not allow us to descend to frequencies below the loudspeaker resonance.\\
	From the frequency-domain transfer functions $H_{loc}$ and $H_{dist}$, the discretized time signals are obtained through zero-order-hold transform, and then translated into the corresponding pressure-dependent current in the loudspeakers coils.\\
	
	\section{Acoustic passivity of the controlled electroacoustic devices}\label{app:passivity_control}
	
\begin{figure}
		\centering
		\includegraphics[width=8.6cm]{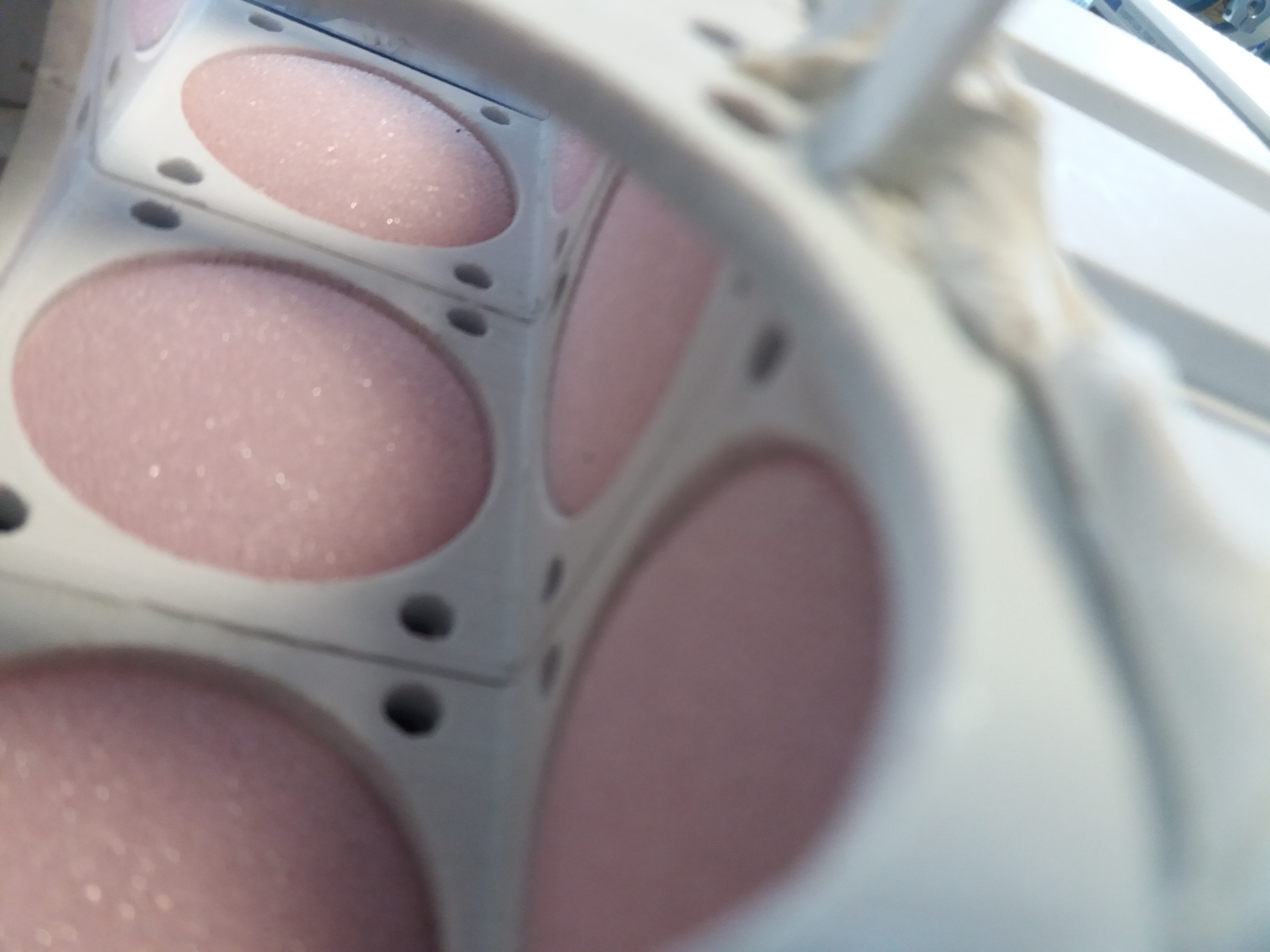}\\
		\caption{View of the waveguide interior. The layers of foam, in pink color applied in front of the loudspeakers, do not interfere with the diaphragms mechanical vibration. The "flush" condition of the liner is maintained.}
		\label{fig:foam_intube}
	\end{figure}

	The acoustic passivity demonstrations in Section \ref{sec:acoustic power analytical} and \ref{sec:acoustic power numerical} suppose the capability to exactly reproduce exactly the boundary control law of Eq. \ref{eq:bndcontrol} and \ref{eq:bndcontrol_discretized}. Clearly, this is not true in a real application, where delay and model incertitudes are inevitably present in the controller. Above all, the time delay is responsible of the loss of acoustic passivity at high frequencies. This can be coped with, by adding some passive acoustic elements in front of the actuators. By applying a thin layer ($6$mm) of melamine foam in front of the loudspeaker (see Fig. \ref{fig:foam_intube}), we restored the passivity at high frequencies (above 3 kHz), untouching the boundary control law performance in the frequency range of interest, where the thin porous layer is practically transparent.
	
	Also, the undesired effect of passivity breaking caused by the delay, and worsened by the model incertitudes, becomes more critical as the gain $g$ adopted in the control transfer functions (see Appendix \ref{app:exp details}) increases. This is the reason why, in this first implementation, $g$ has been kept equal to 0.3.
	
	\section{Equivalent 1D reduced model parameter d}\label{app:equivalent gain}
	
	In order to compare the measurements with the simulations in Section \ref{sec:numerical}, we computed the parameter $d$ relative to an equivalent 1D reduced model of our actual experimental application. To do so, we take into account that the walls of the duct are not fully lined by the actuators, which actually covers just a percentage of the entire duct walls. Therefore, we define an equivalent perimeter $L_p^{eq}$ of the duct, which takes into account the percentage of the actual area covered by each loudspeaker actuator respect to the full area covered by the single cell.
	
	\begin{equation}
	L_p^{eq} = L_p \frac{S_d}{S_{cell}}
	\label{eq:equivalent perimeter}
	\end{equation}
	
	The area of the actuator is taken equal to the effective piston area $S_d$ of the Thiele-Small modelization of the loudspeaker, which in our case is about 0.001$m^2$. The effective area of the cell ($S_{cell}$) instead is given by the lateral dimension of the cell ($L_{cell}=5.5$cm), squared. Therefore:
	
	\begin{equation}
	d_{eq} = g \frac{L_p^{eq}}{S}
	\label{eq:d_eq}
	\end{equation}
	Where $g$ is the gain adopted experimentally in the filter of the synthesized control law, and is equal to 0.3 (see Appendix \ref{app:exp details} and \ref{app:passivity_control}). The parameter $d_{eq}$ of the 1D-reduced model equivalent to the experimental setup, would therefore amount to approximately $d_{eq} = 0.14$ m.\\

	\section*{ACKNOWLEDGEMENTS}
	This project has received funding from the European Union's Horizon 2020 research and innovation programme under the Marie Sklodowska-Curie grant agreement No 722401.\\
	
	\bibliographystyle{unsrt}

	
\end{document}